\documentclass[a4paper,fleqn,usenatbib]{mnras}

\usepackage[T1]{fontenc}
\usepackage{ae,aecompl}
\usepackage{graphicx}
\usepackage{amsmath}
\usepackage{amssymb}
\usepackage{hyperref}

\title[dSphs in MOG]{Testing modified gravity with dwarf spheroidal galaxies }

\author[Haghi \& Amiri]
{Hosein Haghi$^{1}$\thanks{
E-mail:  \mbox{haghi@iasbs.ac.ir} (HH),
 },  \textit{and} Vahid Amiri$^{1}$ \\\\
$^{1}$Department of Physics, Institute for Advanced Studies in Basic Sciences (IASBS), PO Box 11365-9161, Zanjan, Iran\\
}

\begin{document}

\date{Accepted \ldots. Received \ldots; in original form \ldots}

\pagerange{\pageref{firstpage}--\pageref{lastpage}} \pubyear{2016}

\maketitle

\label{firstpage}

\maketitle

\begin{abstract}

The observed velocity dispersion of the classical dwarf spheroidal (dSph) galaxies of the Milky Way (MW) requires the Newtonian stellar mass-to-light ($M_*/L$) ratios in the range of about 10 to more than 100 solar units that are well outside the acceptable limit predicted by stellar population synthesis models. Using Jeans analysis, we calculate the line-of-sight velocity dispersion ($\sigma_{\emph{los}}$) of stars in eight MW dSphs in the context of the modified gravity (MOG) theory of Moffat, assuming a constant $M_*/L$ ratio without invoking the exotic cold dark matter. First, we use the weak field approximation of MOG and assume the two parameters $ \alpha $ and $ \mu $ of the theory to be constant as has already been inferred from fitting to the observed rotational data of The HI Nearby Galaxy Survey catalogue of galaxies. We find that the derived $M_*/L$ ratios for almost all dSphs are too large to be explained by the stellar population values. In order to fit the line-of-sight velocity dispersions of the dSph with reasonable $M_*/L$ values, we must vary $\alpha$ and $\mu$ on a case by case basis. A common pair of values cannot be found for all dSphs. Comparing with the values found from rotation curve fitting, it appears that $\mu$ correlates
strongly with galaxy luminosity, shedding doubt on it as a universal constant.

\end{abstract}


\section{Introduction}

Dwarf spheroidal (dSph) galaxies are among the oldest structures and are by far the most numerous galaxies in the Universe. They populate the very faint end of the galaxy luminosity function.

The Milky Way (MW)  harbours about 19 dSph satellite galaxies \citep{walker07}, distributed from a few tens to a few hundred kpc from the Galactic Centre. The stellar content of the dSphs is typically similar to that of globular clusters, except that perhaps the dSph galaxies have a large amount of dark matter (DM). According to the standard picture of galaxy formation, the dSph galaxies are formed in primordial subhaloes that capture enough gas to form stars,  hence they appear today as very faint, DM-dominated objects (e.g., \citealt{white78,stoehr02}).

Alternatively, Metz and Kroupa have argued that some of the MW's dSphs may be of tidal origin. They found that the tidal dwarf galaxies (TDGs\footnote{TDGs are among the secondary structures that form when gas-rich galaxies interact. Indeed, conservation of angular momentum and energy leads to expanding tidal arms that fragment and form dwarf-galaxy-type objects.}) appear after a Hubble time of dynamical evolution in the host DM halo, as objects that are known as dSph satellite galaxies  \citep{metz07}.

Using Jeans analysis and assuming a specific spherical density profile in the equilibrium state, one can estimate the mass-to-light ratio of the dSph galaxies from the observed data of line-of-sight (los)-velocity dispersions. If one considers the dSph galaxies as DM-dominant objects, then utilizing the Newtonian gravity to explain the velocity dispersion leads to high values of the dynamical mass-to-light ($M/L$) ratios.

Gilmore  considered the dSphs as cold dark matter (CDM)-dominated systems and found that their central dark mass densities should be about $\sim 10^8 M_\odot kpc^{-3} $  with the total DM mass of $ 4 \times 10^7 M_\odot$. He also computed the dynamical $ (M/L)_{V} $ ratios of these systems being about $1.5-320$ solar units \citep{gilmore07a, gilmore07b}.

The dSph galaxies are also in the regime of low acceleration and therefore they can provide a critical test for alternative gravity models such as Modified Newtonian Dynamics of Milgrom  \citep{milgrom83a, milgrom83b} and modified gravity (MOG) of Moffat \citep{moffat06}.

Eight of the MW's dSphs (i.e., Carina, Draco, Fornax, Sculptor, Sextans, Leo I, Leo II, and Ursa Minor) are relatively nearby for which the los-velocity dispersion data as a function of projected radius are well measured \citep{walker07}.  Using Jeans analysis in MOND, Angus (2008) showed that the stellar $M/L$ ratio of four out of eight dSphs with the highest surface brightness (highest internal gravities) are compatible with the stellar population synthesis (SPS) models with reasonable $M/L$ ratios, but the other four required $M/L$ ratios that were larger than the expected range of 1 - 5 in the \emph{V}-band found from stellar population modelling (Maraston 2005).  He argued that the latter four dSphs may be subject to tidal forces that produce tidally unbound interloper stars and inflate the velocities of the bound stars.  Furthermore, using the MOND \emph{N}-body simulations, Angus et al. (2014) found that the range of $M/L$ ratios that best reproduces the observed los-velocity dispersions of Carina is 5.3 - 5.7. They concluded that some tension still exists between the required $M/L$ ratio for the Carina dSph in MOND and those expected from SPS models even by varying of the external acceleration on the more plunging orbits.

In this paper, we aim at studying the internal dynamics of dSphs in the framework of the MOG theory, which has been proposed by Moffat as an alternative to CDM scenarios. It is a fully covariant theory of gravitation that is based on a relativistic action principle involving a vector field of non-zero mass. There have been several attempts at testing MOG in different astrophysical regimes: from globular clusters to cosmological scales. The MOG theory has arguably been successfully used to explain the internal dynamics of globular clusters \citep{toth08}, a large number of galaxy rotation curves \citep{brownstein06a, rahvar13}, kinematics of Magellanic stream \citep{haghi10},  galaxy cluster mass profiles \citep{brownstein06b}, cosmological observations and structure formation\citep{toth13, roshan14}, the CMB acoustical peaks \citep{moffat14}, and gravitational lensing in the bullet cluster IE0657-56 \citep{brownstein07}.

In this paper we examine, for the first time, the MOG theory with available observational data of eight MW dSphs and find their  stellar mass-to-light ratio by solving the Jeans equation and fitting the MOG theory to the dSphs  los-velocity dispersion data. The paper is organized as follows: in Section 2, we briefly review the current status of the MOG theory and introduce the Jeans equation in the weak field limit of MOG;  results and discussion are presented in Section 3; the paper is concluded in Sect. 4.

\section{The Jeans equation in MOG}

\subsection {The MOG theory}

The MOG action is given by
	\begin{equation}
		\label{e1}
		S=S_G + S_\phi +S_S + S_M,
	\end{equation}
where, $S_G$, $S_\phi$, $S_S$, and $S_M$ are Einstein gravity, massive vector field, scalar field, and pressureless dust actions, respectively. The extended form of these actions are as follows:
	\begin{equation}
    	\label{e2}
		S_G=\frac{-1}{16\pi}\int \frac{1}{G} (R+2\Lambda)\sqrt{-g} \: d^{4}x, \\
	\end{equation}
	\begin{eqnarray}
		\label{e3}
		S_\phi=\frac{-1}{4\pi}\int \omega \Bigg[\frac{1}{4}B^{\mu\nu}B_{\mu\nu}&-&\frac{1}{2}\mu^{2} \phi_\mu \phi^{\mu} \nonumber \\
		&+&V_\phi (\phi_\mu \phi^{\mu}) \Bigg]\sqrt{-g} \: d^{4}x,
	\end{eqnarray}

	\begin{eqnarray}
		 \label{e4}
		 S_S=-\int \frac{1}{G} &\Bigg[& \frac{1}{2}g^{\alpha\beta}(\frac{\nabla_{\alpha}G\nabla_{\beta}G}{G^2}+\frac{\nabla_{\alpha}\mu\nabla_{\beta}\mu}{\mu^2}) \nonumber \\
		&+&\frac{V_{G}(G)}{G^2}+\frac{V_{\mu}(\mu)}{\mu^2}\Bigg] \sqrt{-g} \: d^4x,
	\end{eqnarray}
	
	\begin{equation}
	\label{e5}
		S_M=\int(-\rho \sqrt{u^{\mu}u_{\mu}}-\omega Q_5 u^{\mu}\phi_{\mu})\sqrt{-g} \: d^4x,
	\end{equation}
where, $ B_{\mu\nu}=\partial_{\mu}\phi_{\nu}-\partial_{\nu}\phi_{\mu} $  is the Faraday tensor of the vector field, $ \nabla_{\nu} $  is the covariant derivative with respect to the metric $ g_{\mu\nu} $, $ \omega $ is a dimensionless coupling constant, $G$ is the scalar field representing the gravitational coupling strength, $ \mu $ is a scalar field corresponding to the mass of the vector field, $ V_{\phi}(\phi_{\mu}\phi^{\mu}), V_G(G) $, and $V_{\mu}(\mu) $ are the self-interaction potentials associated with the vector field and the scalar field, $\rho$  is the density of matter, and $Q_5=\kappa\rho$ is the fifth force source charge, where $\kappa$ is constant \citep{rahvar13}. It is possible to acquire the exact static spherically symmetric solution of the MOG field equation for a point-like mass  by using the weak field approximation for the dynamics of the fields \citep{toth09}.

Perturbing the fields around Minkowski space--time for an arbitrary distribution of non-relativistic matter, one can obtain effective potential, $Q_{eff}$, and gravitational acceleration, $a(\vec{x})$ as follows:

\begin{equation}
\label{e6}
Q_{eff}(\vec{x})\!=\!-G_N \!\! \left[\!\int\!\!\frac{\rho(\vec{x'})}{\mid\!\vec{x}\!-\!\vec{x'}\!\mid}(1\!+\!\alpha\!-\!\alpha e^{-\mu\mid\vec{x}-\vec{x'}\mid})d^3x' \!\right],\\
\end{equation}

\begin{eqnarray}
		\label{e7}
	    a(\vec{x})=&-&G_N\int\frac{\rho(\vec{x'})(\vec{x}-\vec{x'})}{\mid\vec{x}-\vec{x'}\mid}\Big[1+\alpha \\
		&-&\alpha e^ {-\mu \mid \vec{x}- \vec{x'} \mid}(1+\mu\mid\vec{x}-\vec{x'}\mid)\Big]d^3x' . \nonumber
 \end{eqnarray}

In the weak--field approximation, $ \alpha $  and $ \mu $  are constants, depending on the mass of the source, in the exact static spherically symmetry solution \citep{toth09}. Acceleration law of a point-like mass in the weak gravitational field regime of MOG contains a Yukawa-shape force term added  to Newtonian acceleration term  as follows \citep{moffat06}:\\
\begin{equation}
\label{e8}
a_{MOG}(\vec{r})=\frac{G_NM}{r^2} \lbrace{1+\alpha[1-e^{-\mu r}(1+\mu r)]\rbrace}.\\
\end{equation}

Here, $G_N$  is the Newtonian gravitational constant,  $\alpha$ is determined by the coupling of the strength of the fifth force vector, $\phi_\mu$,  to the baryon matter, $\mu$ is the range of the force, and $M$ is the total baryonic mass within the radius $r$.

It should be noted that in MOND, the so-called  external field effect (EFE) must be taken into account when dealing with objects embedded in the external tidal field of a more massive system \citep{angus08, haghi09, Malekjani09, haghi11, haghi16}.  This is because of the intrinsic non-linearity of the Poisson equation that is related to the violation of the strong equivalence principle in MOND.  Although the weak equivalence principle is also violated in MOG due to a direct coupling between matter and the Proca vector field, MOG leads to the linear differential equations in the low curvature limit. That is, two linear differential equations are replaced by the standard Poisson equation. In other words,  the internal dynamics of a non-isolated stellar system can be decoupled from the host environment in MOG as in the Newtonian gravity such that a uniform external field does not affect the internal dynamics of a stellar system.

\begin{figure}
\includegraphics[width=85mm]{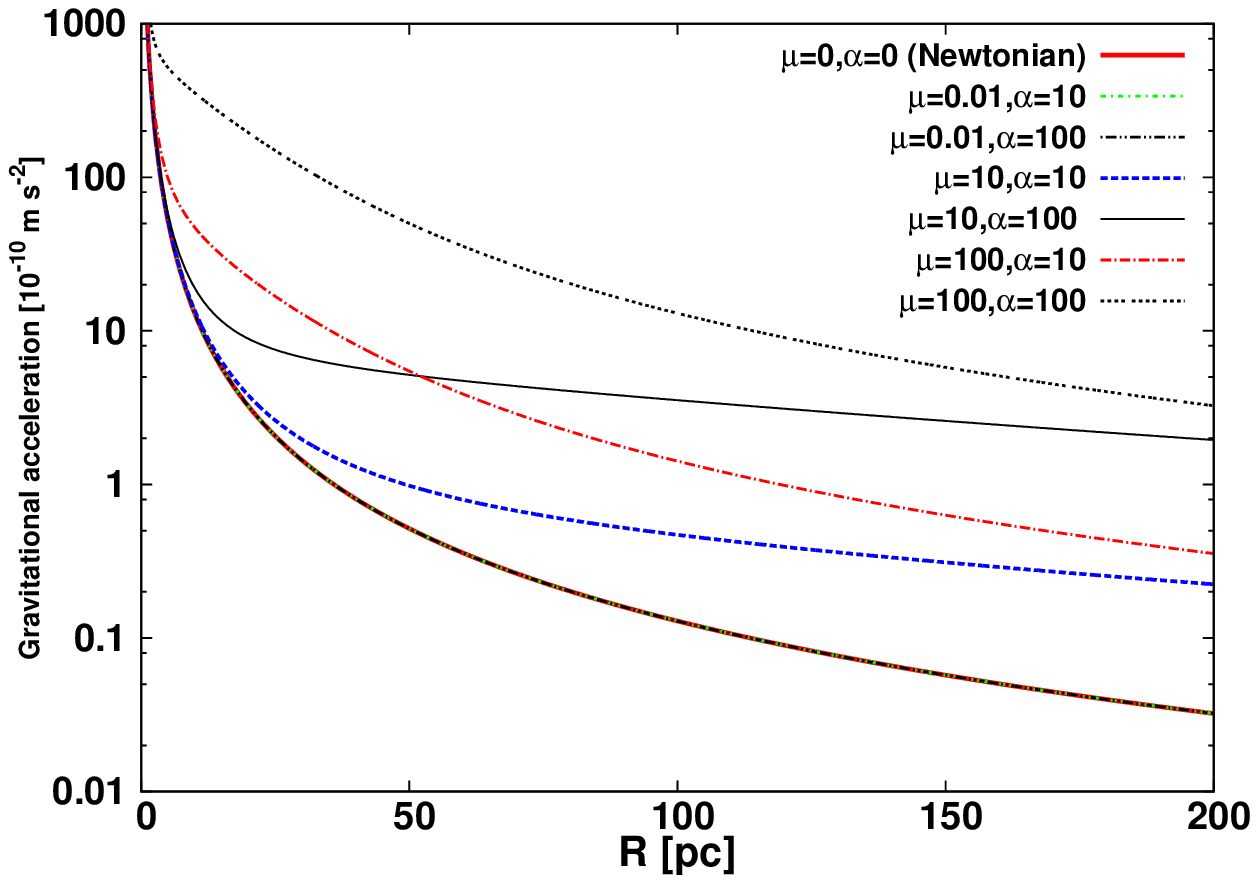}
\includegraphics[width=85mm]{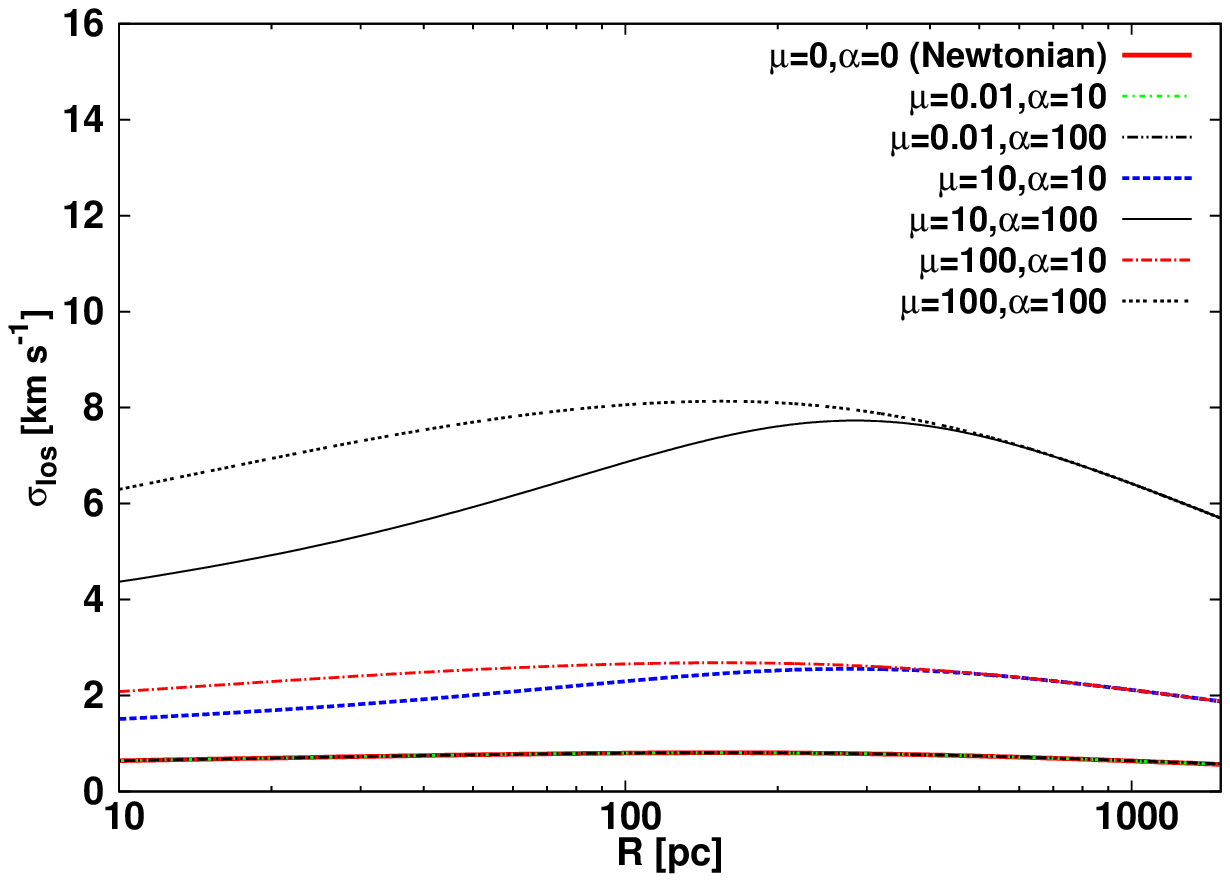}
\caption{Gravitational acceleration (upper panel) and los-velocity dispersion (lower panel) in MOG theory as a function of galactocentric distance for a typical galaxy with the total mass of  $10^6 M_{\odot}$, for different values of $\alpha$ and $\mu$. The gravitational acceleration in MOG theory is very close to the Newtonian gravitational acceleration for smaller values of $\mu$. The case of $\mu=0.01$ is completely Newtonian such that the corresponding curve is hidden behind the Newtonian curve. For the same values of $\alpha$,  the acceleration curves drop rapidly for the higher value of $\mu$ in far distances from the centre of galaxy. See Sec. 3.1 for further details. }
\label{f2}
\end{figure}

\subsection{Jeans analysis}
We  use the Jeans equation to find the radial velocity dispersion in a spherically symmetric gravitational field \citep{binney08}: \\
\begin{equation}
\label{e9}
  \frac{d(\nu(r) \sigma_{r}^2(r))}{dr}+\frac{2\nu(r)}{r} \beta(r) \sigma_{r}^2(r)=-\nu(r) \frac{d\phi}{dr},
  \end{equation}
where, $ r $, $\nu(r)$, $ \beta(r)$, $\sigma_{r}^2(r) $,   and $\phi(r) $  are the radial distance from the dSph centre, the spatial number density of stars in the dSph, the velocity anisotropy, the radial velocity dispersion as a function of radial distance, and the gravitational potential, respectively. One can write

\begin{equation}
\label{e10}
 \frac{d \sigma_{r}^2(r)}{dr} + \frac{A(r) \sigma_{r}^2(r)}{r}=-a(r),
  \end{equation}
where, $A(r)=2\beta(r)+\gamma(r) $ and $ \gamma(r)=\mathrm{d} \:\ln\nu(r)/\mathrm{d}\: \ln r $. Generally $ \beta $ could be zero (i.e., isotropic velocity dispersion), constant, or a function of radial distance from the centre of cluster \citep{toth08}.

The  los-velocity dispersion which is the directly measurable dSph property  as a function of projected distance from the centre of galaxy is given by

\begin{equation}
\label{e11}
\sigma^2_{\textit{los}}(R)= \frac{\int_{0}^{\infty} [y^2+(1-\beta(r))R^2]r^{-2} \sigma_{r}^2(y) \nu (y) dy}{\int_{0}^{\infty} \nu (y) dy},
\end{equation}
where $ y=\sqrt{r^2-R^2} $.  $R$  and $r$ are projected and 3D distances from the centre of the dSph, respectively.  Substituting $y$ into Eq. 11, $\sigma_{\textit{los}}(R)$ can be written as

\begin{equation}
\label{e12}
  \sigma^2_{\textit{los}}(R)\!=\!\frac{\int_{R}^\infty (r^2\!-\!\beta(r)R^2)\sigma_{r}^2(r) \nu(r)/r\sqrt{r^2\!-\!R^2}\: dy}{\int_{R}^\infty r \nu(r)/\sqrt{r^2-R^2} \:dy }.
 \end{equation}

In this paper, the 2D exponential luminosity profile, $L=L_0\exp(-r/r_e)$, is used for the enclosed luminosity profile. The best-fitting values of $r_e$ for each dSph are taken from \cite{Irwin95} and are listed in Table 1. The constant values of $L_0$ for each individual dSph is calculated by assuming that the integration of the luminosity profile from zero to infinity must be equal to the total luminosity, $L_V$ (Table 1).  Assuming a spherical symmetry, the fitted 2D surface luminosity distributions, $I(R)$ can be deprojected into numerical 3D luminosity density distributions, $j(r)$ using an Abel transformation as \citep{binney08}

\begin{equation}
\label{e13}
  j(r)=-\frac{1}{\pi}\int^{\infty}_{r}\frac{dI}{dR}\frac{dR}{\sqrt{R^2-r^2}}.
 \end{equation}

The enclosed luminosity profiles of dSphs are used to numerically calculate the luminosity density as a function of 3D radius for solving the Jeans equation for the radial velocity dispersions.

\begin{figure*}
\includegraphics[width=148mm]{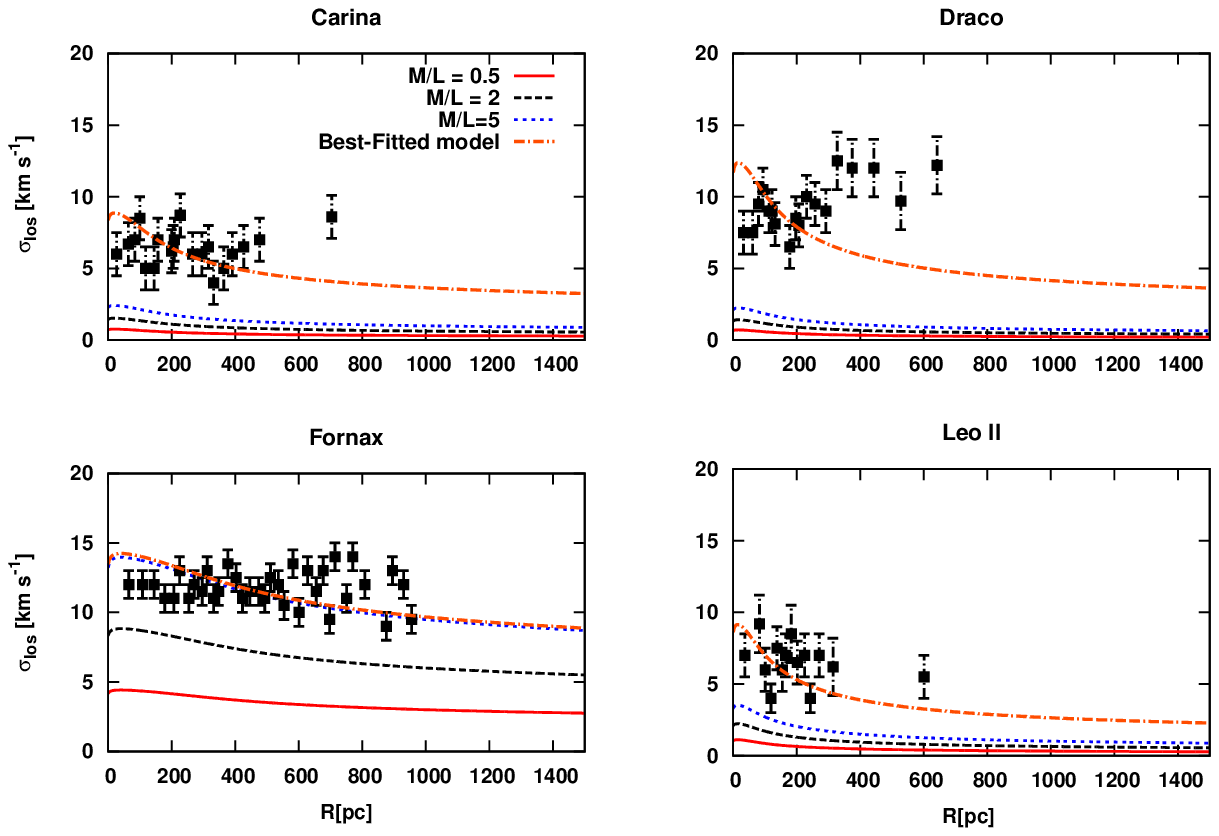}
\includegraphics[width=148mm]{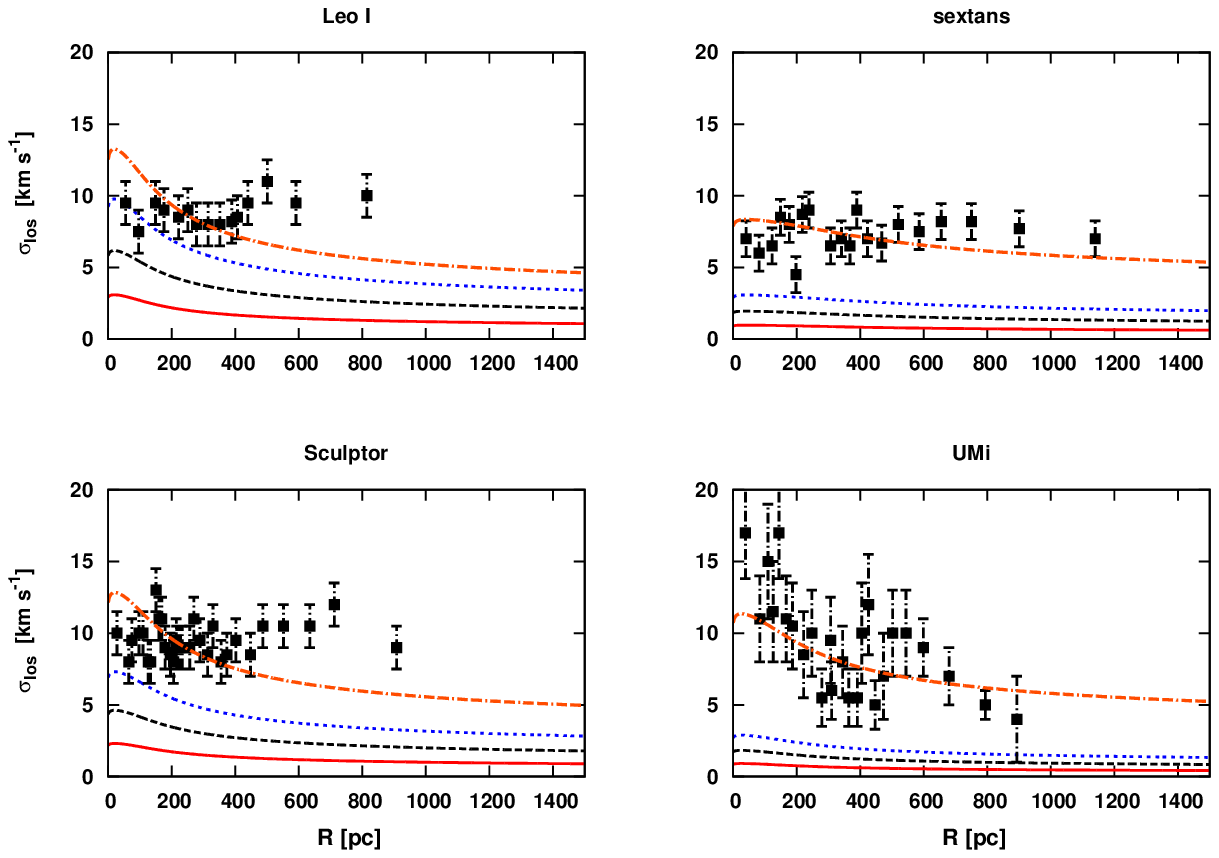}
\caption{los-velocity dispersion profiles as a function of projected distance, for all eight dSphs obtained from the Jeans equation in MOG using the constant values of $ \alpha_{RC} $ and $ \mu_{RC} $  for the MOG parameters. Different lines show the \textit{los}-velocity dispersion assuming different values of stellar mass-to-light ratios. Observational data points (black symbols with error bars) are taken from Walker et al. (2007).  The corresponding $\chi^2$ values are given in Table 1. The large values of $\chi^2$ show that the MOG theory with constant values of $\alpha _{RC}$ and $\mu_{RC}$  could not produce an acceptable fit to the observational data of the MW dSphs, preventing one to conclude $\alpha$ and $\mu$ as the universal parameters. }
\label{f1}
\end{figure*}

\section{Results} \label{S3}

In this section, we present the results of MOG fits to the observed los-velocity dispersion data of MW dSphs. First, we assess the general effect of the MOG parameters (i.e., $\alpha$ and $ \mu $)  on the shape of the velocity profile of a typical dSph galaxy in the MOG dynamics.

\subsection{Velocity dispersion: The General Case}

According to equation (\ref{e8}), at the limit of  $ \mu \ll 1$, the MOG acceleration $(a_{MOG})$ is nearly independent of  $ \alpha$ and $\mu $. At the other limit, in the regime of $ \mu \gg 1 $, the MOG acceleration becomes independent of $ \mu $. Therefore, one can obtain the following relations between MOG  and Newtonian acceleration in the weak filed limit:

\begin{eqnarray}
\mu \ll 1  \Longrightarrow  \mu r \longrightarrow 0, \quad  e^{-\mu r} \longrightarrow 1,\quad a_{MOG}=a_N
\nonumber
\end{eqnarray}

\begin{eqnarray}
\mu \gg 1  \Longrightarrow    e^{-\mu r} \longrightarrow 0, \quad  a_{MOG}=a_N(\alpha +1) .
\nonumber
\end{eqnarray}

In order to investigate the general effect of MOG parameters on the shape of a typical velocity dispersion curve, we calculate the gravitational acceleration of a galaxy that consists of a stellar component with the total mass of $M=10^6 M_{\odot}$ (see the upper panel of Fig. 1). In order to cover the different values of $\alpha$ and $ \mu $, we change the parameters in the range $\alpha=$10--100, and $ \mu= $ 0.01--100 kpc$^{-1}$.

There is a remarkable difference in the predicted velocity dispersion of the inner part of a dSph galaxy in the context of MOG with different values of $\alpha$ and $ \mu $ as shown in Fig. \ref{f2}. In all curves, the velocity dispersion profiles predicted by MOG lie above the Newtonian one. As $\alpha$ and $ \mu $  get smaller, the velocity dispersion curve shifts to the Newtonian prediction.  In particular, for $\alpha \gg1$ and $ \mu \gg 1$ the velocity curve shows a flat behaviour in the outer parts. Especially,  the $\mu$-parameter has a significant effect on the velocity dispersion at the outskirts of galaxies.

\subsection{\texorpdfstring{$\alpha$ and $\mu$ as constants}{alpha and mu as constants}}

By applying MOG to the observed rotation curves of The HI Nearby Galaxy Survey (THINGS) catalogue of galaxies, Moffat and Rahvar  (2013) found the best-fitting values of $\bar{\alpha}_{RC} = 8.9 \pm 3.4 $ and $ \bar{\mu}_{RC} = 0.042 \pm 0.004$ kpc$^{-1} $. Assuming these parameters as universal constants, they also applied the MOG effective gravitational potential to the Ursa-Major catalogue of galaxies and showed that the MOG theory can successfully reproduce the observed data of rotational velocities \citep{rahvar13}.

We use the formalism described in Section 2 to calculate the predicted los-velocity dispersion  for each of the  eight dSphs of the MW, using the photometric data in Table 1. Given the MOG parameters $ \alpha_{RC}=13.0 $ and $ \mu_{RC}=0.15$ kpc$^{-1}$ (as upper limits that are obtained from applying the MOG theory to galaxy rotation curves), we apply the MOG theory to the los-velocity dispersion of dSphs galaxies.  As  shown in Table \ref{t1}, although $ \chi_{MOG}^2 <\chi_{N}^2 $ for most dSphs,  the inferred  stellar mass-to-light ratios are larger than those  expected from SPS models.

For $ $M/L$ = 0.5$ and $5.0$ solar units (as the minimum and maximum acceptable values from SPS models), and using the  constant values of $ \alpha_ {RC}
$ and $\mu_{RC} $, the MOG theory does not fit the los-velocity dispersion data of dSphs as it is clear from the large $\chi^2$-values shown in Fig. \ref{f1}.

\begin{table*}
\begin{center}
\begin{tabular}{ccccccccccccc}
\hline
1&2&3&4&5&6&7&8&9&10&11\\
\hline
Name & $R_h$ & $L_{V}$ & $L_0$ & $r_e$ & $(M_\star/L_V)_N$ & $\chi^2_N $&$ (M_\star/L_V)_{MOG}$ & $\chi^2_{MOG}$ & $\chi^2_{MOG}$& $\chi^2_{MOG}$ &\\
& (pc)& ( $10^{5} {L_{\odot}}$) & ($L_{\odot} /pc^2$) & (pc)& & & & & $M_\star/L_V=0.5$ & $M_\star/L_V=5$\\
\hline
Carina   & $241$ & $4.4$  & $2.2$ & $161.5$ & $79.0_{-31.9}^{+39.9}$  & $2.0$ & $67.1^{+33.6}_{-26.9}$ &$ 1.6 $&$18.6$ &$12.3$ \\
\hline
Draco   &$ 196 $  & $3.3$ &$ 2.2 $  & $121.7$ & $166.3_{-53.8}^{+64.0}$  & $5.3$ &$152.3^{+78.8}_{-62.2}$ &$ 4.6$ &$34.5$ &$27.1$\\
\hline
Fornax  &$ 668 $&$ 158.0 $ &$ 15.7 $&$ 397.4$& $8.8_{-1.4}^{+1.6}$     & $ 6.2 $& $5.2^{+1.0}_{-0.9}$ &$ 3.1$ &$72.5$ &$3.2$ &  \\
\hline
Leo I   & $ 246 $&$ 76.0 $ & $ 30.9 $&$ 149.5 $& $10.9_{-3.4}^{+4.0}$     & $4.2$ &$9.2^{+3.3}_{-2.8}$ &$ 3.0$&$25.4$ &$5.6$ \\
\hline
Leo II  &  $ 151 $&$ 7.6 $ & $ 7.8 $&$ 101.6 $& $36.5_{-14.5}^{+18.0}$     & $1.9$ &$33.7^{+19.8}_{-11.1}$ &$ 1.8$ &$16.9$ &$9.2$\\
\hline
Sextans &  $ 682 $&$ 6.3 $ & $ 0.7 $&$ 430.5 $&$59.3_{-18.2}^{+21.6}$  & $2.4$ & $36.7^{+13.0}_{-11.2}$&$ 1.5$ &$31.1$ &$16.6$\\
\hline
Sculptor  &  $ 260 $&$ 28.0 $ & $ 14.2 $&$ 172.0 $& $18.4_{-5.4}^{+6.2}$   & $3.6$ & $15.4^{+5.1}_{-4.5}$ &$ 2.6$ &$30.4$ &$10.2$\\
\hline
 UMi   & $ 280 $ & $ 11.0 $  &$ 1.5 $&$ 223.3 $& $113.0_{-55.9}^{+74.8}$    & $0.97$ & $77.0^{+51.2}_{-38.3}$ &$ 1.0$ &$11.4$ &$7.9$\\
\hline

\end{tabular}
\end{center}
\caption{Best-fitting reduced $ \chi^2 $ and stellar $M_*/L_V$ values of eight MW dSph galaxies in the Newtonian and MOG dynamics obtained by fitting MOG theory to the los-velocity dispersion data of dSphs with  $\alpha_{RC} = 13 $ and $ \mu_{RC} = 0.15$ kpc$^{-1}$ as constants.  Column 1  gives the  name of dSph galaxies. Columns 2 and 3 give the  half-mass radius of dSph galaxies from Walker et al. (2009) and the  \emph{V}-band luminosity, respectively. The best-fitting parameters of the exponential luminosity profile for each dSph are given in columns 4 and 5. Columns 6 and 7 contain the best-fitting Newtonian $M_*/L_V$ and corresponding minimum $ \chi^2 $, respectively. Columns 8 and 9 are the best-fitting reduced $ \chi^2 $ and stellar $M_*/L_V$ values in the MOG context with only one free parameter of the stellar mass-to-light ratio.  The $ \chi^2 $ associated with the fit to the MOG theory assuming  constant values of 0.5 and 5 for the stellar  mass-to-light ratios are presented in columns 10 and 11, respectively. }
\label{t1}
\end{table*}
\begin{figure*}
\includegraphics[width=150mm]{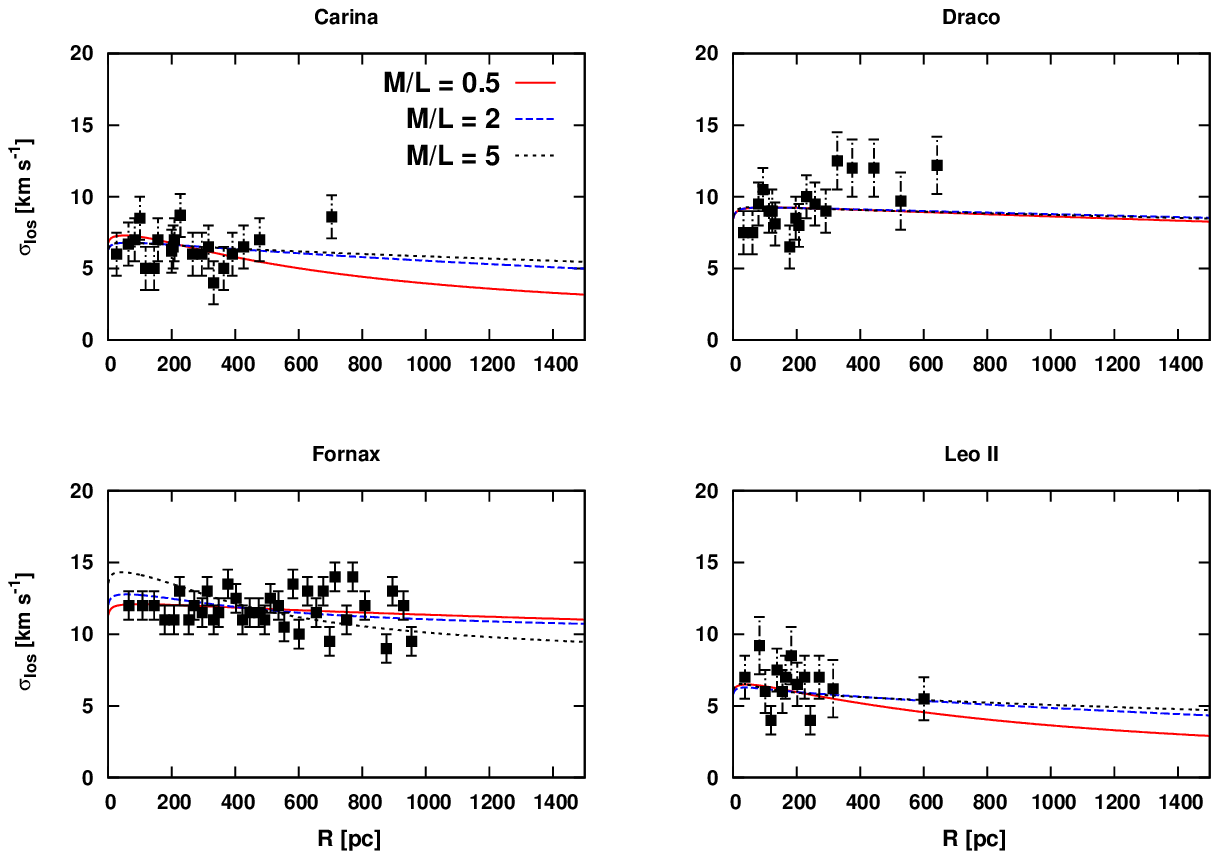}
\includegraphics[width=150mm]{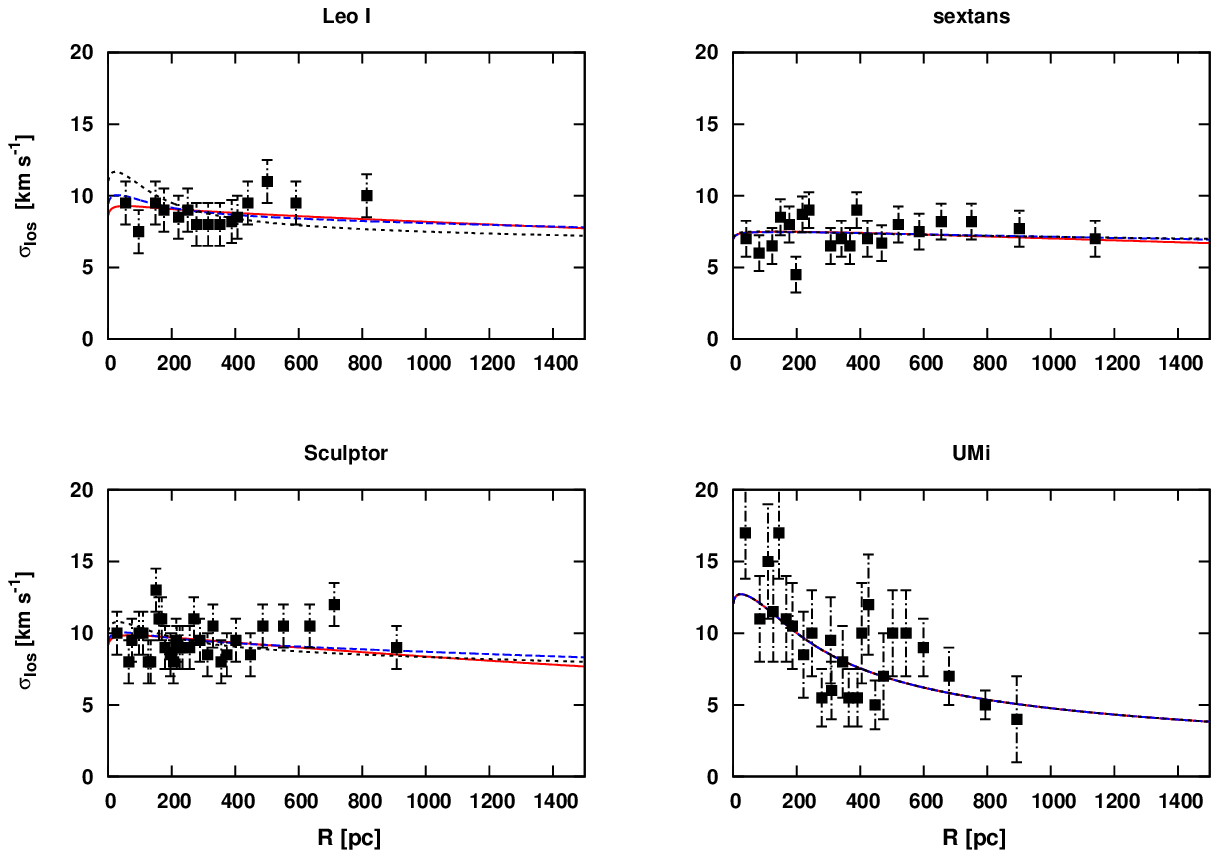}
\caption{Fitting the los-velocity dispersion profiles as a function of projected distance obtained from the Jeans equation to eight MW  dSphs data, letting  $ \alpha$ and $ \mu$ as free parameters of MOG. The best-fitting values of fitting parameters are displayed in Table 2. Observational data points are taken from Walker et al. (2007).  Best-fitting lines  for different values of  stellar $M_*/L_V$ ratios are overplotted.}
\label{f3}
 \end{figure*}

\subsection{\texorpdfstring{$\alpha$ and $ \mu $ as fit parameters}{alpha and mu as fit parameters}}

So far, we have shown that the MOG theory with the constant values of $\alpha$ and $\mu$,  does not give acceptable fits to the los-velocity dispersion profile of dSphs, if we limit the stellar $M_*/L$ ratios to vary in an acceptable range as implied by SPS models.   That is, in order to produce  higher values of gravitational acceleration, and as a result, higher values of los-velocity dispersion; one needs to assign larger values of $\alpha$ and $\mu$ compared to  $\alpha_{RC}$ and $\mu_{RC}$.

According to the classical version of MOG  \citep{ moffat06, toth09}, here in this section we assume  $\alpha$ and $ \mu $ as two scaling parameters that vary with the scalelength and mass of the gravitational source. These parameters  determine the coupling strength of the vector field to baryonic matter and the range of the force. Moffat  argued that  these parameters are scale dependent \citep{moffat06}; thus, they are not to be taken as universal constants. Indeed, these parameters can be determined by the equations of motion for effective scalar fields derived from an action principle.  An empirical fitting of $\alpha$ versus $ \mu $ for the wide range of spherically symmetric systems, from  the Solar system scale to clusters of galaxies has been obtained and depicted in fig. 2 of \cite{brownstein07}. See, e.g., Haghi and Rahvar (2010) for a brief review on this issue.

Fitting the calculated los-velocity dispersion to the observed data points is achieved by adjusting $\alpha$ and $ \mu $ as fitting free parameters, through a $\chi^2$  goodness-of-fit test, defined as

\begin{equation}
 \chi^2=\frac{1}{(N-P-1)}\sum_{i=1}^{N}\frac{(\sigma^{i}_{MOG}-\sigma_{obs}^{i})^2}{\sigma_i^2}\label{chi},
\end{equation}

\noindent where $\sigma_i$ is the observational uncertainty in the los-velocity dispersion and $P$ is the number of degrees of freedom.

In order to show the sensitivity of the results to the assumed stellar mass-to-light ratios, we calculate los-velocity dispersion profile for three values of $M/L_V$=0.5, 2, and 5, which span a plausible range in agreement with the SPS models.

Using an exponential model for 2D  number density profile for the stars in dSph galaxies with isotropic velocity dispersion, we obtained very good fits to the los-velocity dispersion data as a function of projected distance (Fig. 3). Almost all models trace the observed data with reasonable detail. The best-fitting values of $\alpha$ and $ \mu $ as  fitting free parameters, and the minimum $\chi^2$ values are listed in Table 2.

\begin{figure}
	\includegraphics[width=85mm]{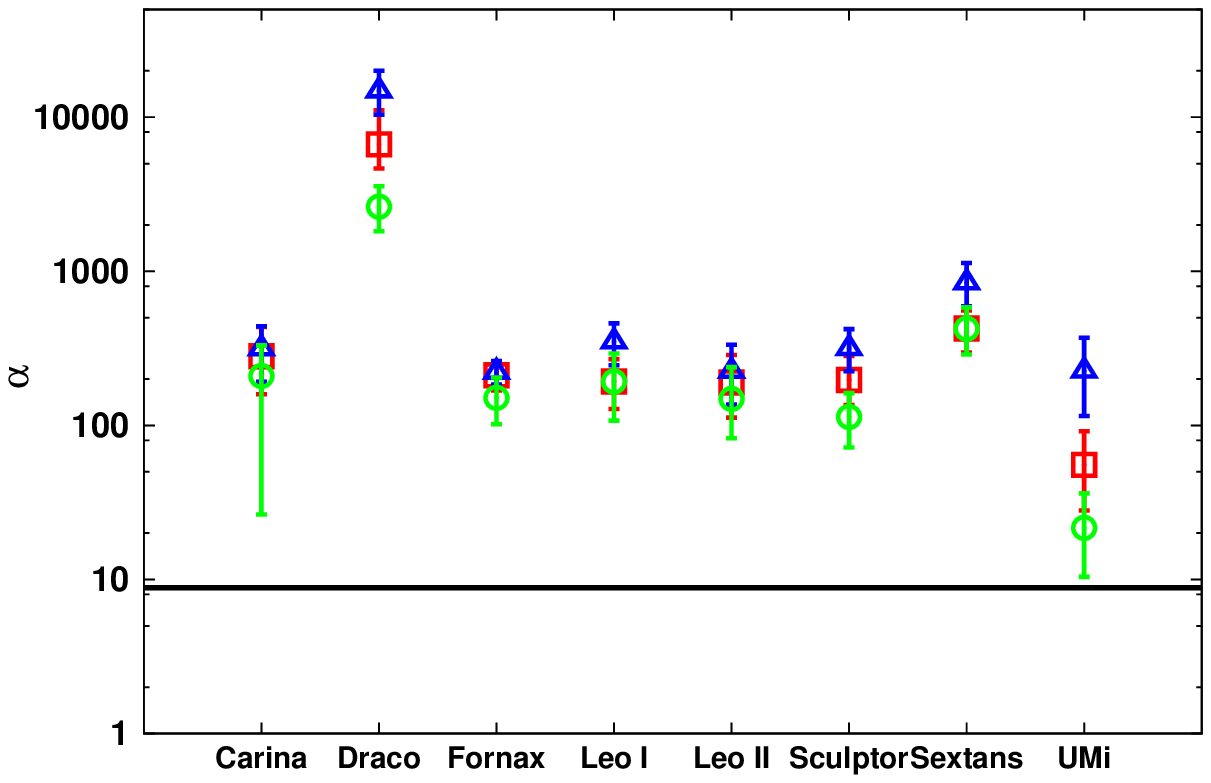}
\includegraphics[width=85mm]{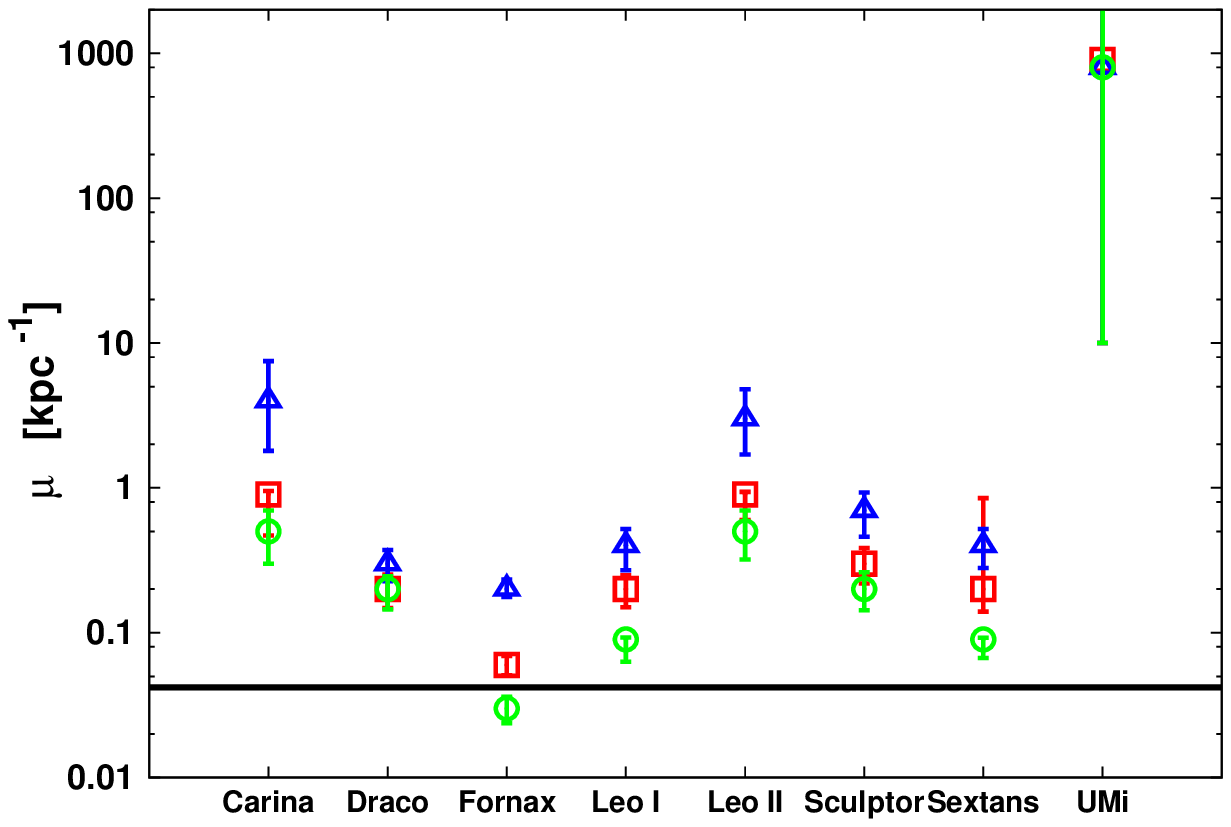}
\caption{  The best-fitting MOG parameters $\alpha$ (top panel) and $\mu$ (bottom panel) with their uncertainties shown as data points.  The solid lines show the  $\alpha_{RC}$ and $\mu_{RC}$ which are the upper limit values that have been inferred from fitting to the observed rotational data of THINGS catalogue of galaxies (Moffat \& Rahvar 2013). Three symbols for each dSph indicate the corresponding values of $\alpha$ and $\mu$ obtained from best-fitting results for $M/L_V=0.5$ (triangle), $M/L_V=2$ (square), and $M/L_V=5$ (circle).}
	\label{f2.5}
\end{figure}

 \begin{table}
\begin{tabular}{cccccc}

 Name& $(M_\star/L_V)$ & $\alpha$ & $\mu$ (kpc)$^{-1}$ & $\chi^2$\\
 \hline
              & $0.5 $ &$ 314.7^{+125.2}_{-122.9}$ &$ 4^{+2.2}_{-3.5}$&$ 1.12$\\
 Carina       & $2.0$  & $281.4^{+155.1}_{-122.2}$ &$ 0.9^{+0.5}_{-0.4}$&$ 0.87$\\
              & $5.0$  &$ 209.2^{+121.3}_{-94.9} $  & $0.5^{+0.2}_{-0.2}$&$ 1.03$\\
\hline
              &$ 0.5$  &$ 14794^{+5206}_{-4407}$ &$ 0.3^{+0.07}_{-0.07}$&$ 1.22$\\
    Draco     &$ 2.0$  &$6660^{+2370}_{-2014}$ &$ 0.2^{+0.05}_{-0.05}$&$ 1.22$\\
              &$ 5.0 $ &$ 2623^{+942}_{-803} $  & $0.2^{+0.04}_{-0.05}$&$ 0.81$\\
\hline
              &$ 0.5 $ &$ 221.2^{+40.3}_{-37.1}$ &$ 0.2^{+0.03}_{-0.02}$&$ 1.63$\\
 Fornax       & $2.0 $ & $211.4^{+46.1}_{-42.7}$ &$ 0.06^{+0.009}_{-0.009}$&$ 1.81$\\
              & $5.0$  &$ 150.8^{+53.4}_{-49.1} $  & $0.03^{+0.006}_{-0.006}$&$ 2.97$\\
\hline
              &$ 0.5$  &$ 348.8^{+120.0}_{-102.7}$ &$ 0.4^{+0.1}_{-0.1}$&$ 0.58$\\
    Leo I     &$ 2.0$  & $192.7^{+76.4}_{-64.7}$ &$ 0.2^{+0.05}_{-0.05}$&$ 0.71$\\
              &$ 5.0$  &$ 192.6^{+99.9}_{-85.1} $  & $0.09^{+0.02}_{-0.02}$&$ 1.31$\\
\hline
              &$ 0.5$  &$ 224.6^{+109.4}_{-88.1}$ &$ 3^{+1.8}_{-1.3}$&$ 1.38$\\
     Leo II   & $2.0 $ & $190.1^{+96.2}_{-77.8}$ &$ 0.9^{+0.3}_{-0.3}$&$ 1.34$\\
              &$ 5.0$  &$ 148.7^{+90.2}_{-66.2} $  & $0.5^{+0.2}_{-0.1}$&$ 1.32$\\
\hline
              &$ 0.5$  &$ 842.0^{+290.5}_{-248.5}$ &$ 0.4^{+0.1}_{-0.1}$&$ 0.94$\\
    Sextans   &$ 2.0$  & $426.0^{+131.5}_{-128.9}$ &$ 0.2^{+0.06}_{-0.06}$&$ 0.93$\\
              & $5.0$  &$ 424.3^{+159.0}_{-135.3} $  & $0.009^{+0.02}_{-0.02}$&$ 0.94$\\
\hline
              &$ 0.5$  &$ 315.1^{+106.5}_{-90.6}$ &$ 0.7^{+0.2}_{-0.2}$&$ 0.86$\\
  Sculptor    & $2.0$  & $197.2^{+89.4}_{-57.2}$ &$ 0.3^{+0.08}_{-0.08}$&$ 0.84$\\
              & $5.0 $ &$ 113.4^{+48.5}_{-41.6} $  & $0.2^{+0.06}_{-0.05}$&$ 1.01$\\
\hline
              &$ 0.5$  &$ 224.9^{+145.6}_{-109.8}$ &$ 800.0^{+\infty}_{-790.0}$&$ 1.01$\\
    UMi       & $2.0 $ & $55.5^{+36.6}_{-27.5}$ &$ 900.0^{+\infty}_{-890.0}$&$ 1.01$\\
              &$ 5.0$  &$ 21.6^{+14.6}_{-11.2} $  & $ 800.0^{+\infty}_{-780.0}$&$ 1.68$\\

\end{tabular}
\caption{\label{t2} Best-fitting reduced $ \chi^2 $ and stellar $M_*/L_V$ values of eight MW dSph galaxies obtained from the Jeans equation in the MOG gravity, assuming $\alpha$ and $ \mu $  as fitting parameters.  Column~1 gives the galaxy name. The second column displays three different values for stellar mass-to-light ratios in \emph{V}-band used in our analysis which are justifiable by SPS models. The following three columns give the best-fitting values of free parameters and the corresponding minimum $ \chi^2 $ values.  The average values of $ \alpha $ and $ \mu $ are $221\pm 112$ (excluding the case of Draco) and  $0.41\pm0.35$ kpc$^{-1}$ (excluding the case of UMi), respectively. The best-fitting values in the case of $M_*/L_V=2$ is used to calculate the average values of $ \alpha $ and $ \mu $.}
\end{table}

\begin{figure}
	\includegraphics[width=85mm]{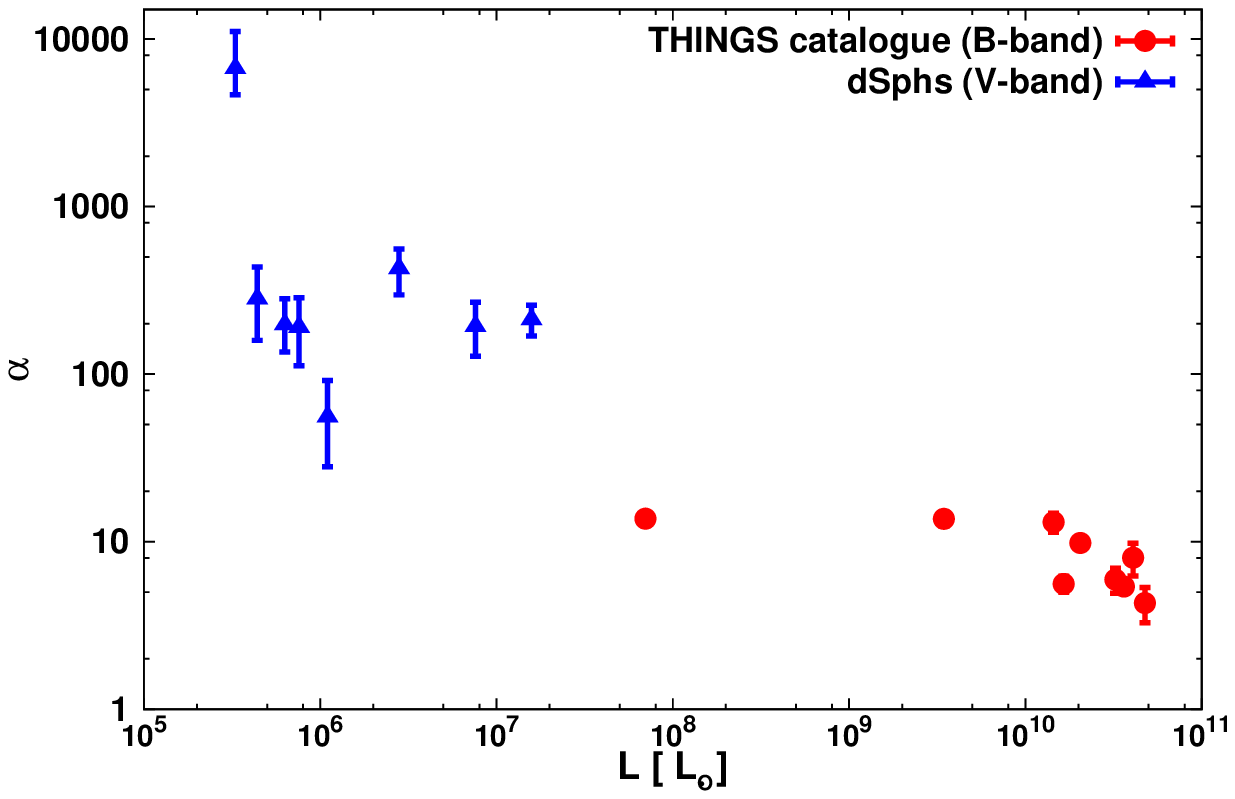}
\includegraphics[width=85mm]{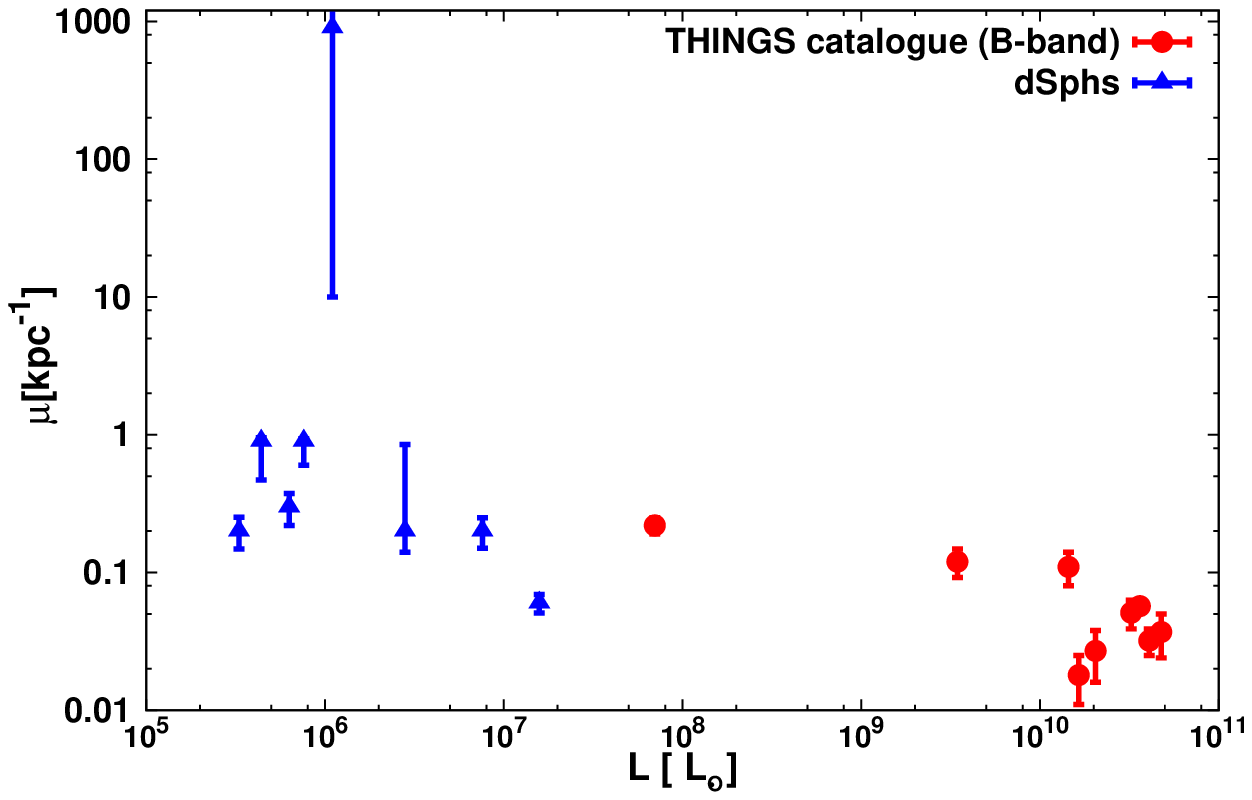}
\caption{  The best-fitting MOG parameters $\alpha$ (top panel) and $\mu$ (bottom panel) for the case of $M_*/L_V=2$ as given in Table 2 versus the total luminosity of dSphs in \emph{V}-band (triangles) and THINGS catalog of galaxies in \emph{B}-band (circles, Moffat \& Rahvar 2013). Especially, apparent is the fact that the inferred MOG parameters are nearly related to the total luminosity of each individual galaxy. }
\label{f2.52}
\end{figure}


\section{Summary and discussion}

Note that in the MW dSph galaxies which are smaller than MW-like galaxies by a factor of $\sim 50 $ in scale, the average values of  $\alpha$ and $ \mu $ from fitting MOG to los-velocity dispersion data are  $221\pm112$ (excluding the case of Draco) and $0.41\pm0.35$ kpc$^{-1}$ (excluding the case of UMi), respectively. These values are significantly larger than $\alpha_{RC}$ and $ \mu_{RC} $ inferred from the best fits to the rotation curve of spiral galaxies. In Fig. 4, we compare  the best-fitting values of MOG parameters assuming $M/L_V=$ 0.5, 2, and 5 indicated by circles, squares, and triangles, respectively. As can be seen, the $\alpha$-values are nearly distributed  above the value that is inferred from the best fits to the rotation curve of spiral galaxies (i.e., $\alpha_{RC}$). The implied $\mu$-values (in seven out of eightdSph galaxies) are significantly larger  than what is imposed by rotation curve fits (with a scatter between 1 and 2 orders of magnitude), supporting the hypothesis that the MOG parameters are scale dependent. This poses a serious challenge to the universality of MOG since $\alpha$ and $ \mu $ should be the same for every galaxy.

Fig. 5 shows the best-fitting $ \alpha $ and $ \mu $ parameters (assuming $M/L_V=$ 2 ) versus the total luminosity of dSph galaxies. The best-fitting values from fitting to the rotation curves of THINGS catalogue (Moffat \& Rahvar 2013) is also overplotted. As can be seen, the $ \alpha $ and $ \mu $ parameters show a clear correlation with the luminosities of galaxies, so that by increasing the luminosities, the inferred values  of $ \alpha $ and $ \mu $ decrease. However, this is in agreement with the general trend that already have been argued by Brownstein \& Moffat (2007) showing that the best-fitting values of $ \mu $ decrease by moving from the low-mass systems (e.g., dwarf galaxies) to the massive systems (e.g., normal and dwarf X-ray clusters), any systematic trend of $ \alpha $ and $ \mu $ with some galaxy parameters could be a problem for MOG theory since these parameters are supposed to be universal constants.

As we have previously shown in Fig. 1, the gravitational acceleration in MOG is very close to the Newtonian acceleration for smaller values of $ \alpha $ and $ \mu $, while it is stronger than the Newtonian prediction for the larger values of $ \alpha $ and $ \mu $. Therefore, the above anticorrelation between the MOG parameters and the total luminosity of galaxies can be interpreted in favour of the mass discrepancy acceleration relation which is a strict prediction of MOND, and has subsequently been observed for a sample of spiral galaxies (Sanders 1990; McGaugh 2004; Wu \& Kroupa 2015).

We should point out that the main caveat of our calculation is that all dSphs are assumed to be in equilibrium state by adopting a simplified density profile without any tidal interaction with our Galaxy. Indeed the evolution of dSph galaxies in a triaxial time-dependent galactic potential should be considered.  This will be a subject of future investigations.

\section{Conclusion }\label{S4}

As an alternative approach to the CDM hypothesis, covariant MOG theory, which is a scalar--tensor--vector theory, has successfully  explained the dynamics of a wide range of astrophysical systems without invoking any exotic matter. It contains two free parameters $ \alpha $ and $ \mu $  which were thought to be universal parameters. In this study, we used the photometric data to calculate the los-velocity dispersion profiles of eight of the MW dSph satellite galaxies in the framework of the MOG theory based on Jeans analysis and compared with observed kinematical data.

First of all, we used the MOG effective gravitational potential with the universal parameters $ \alpha_{RC}= 13$ and $\mu_{RC}= 0.15$ kpc$^{-1}$ (which are the upper limits fixed from fitting the MOG theory to galaxy rotation curves) to fit MOG to the observed data of velocity dispersion as a function of radius with only one free parameter; the stellar $M_*/L_V$ ratio. Assuming two different profiles for the stellar density of dSphs, we found that the constant MOG parameter models provide a reasonably good fit (with an average reduced $\chi^2\simeq 2.4 $) to the kinematic data of eight dSphs, but the required high $M_*/L_V$ ratios being about 5--150 solar units in \emph{V}-band, are completely inconsistent with the stellar populations, with the exception of the Leo I dSph galaxy.

In addition, if we limit the stellar $M_*/L_V$ ratio of the models to vary within the plausible range (i.e., $M_*/L_V\leq 5$), the velocity dispersion data will be poorly  fitted by MOG.  It should be noted that this conclusion is not affected by changing the stellar density distribution. That is, it is not possible in MOG to describe the dynamics of eight dSph galaxies with a consistent set of parameters.

In order to find better fits, we finally let the two parameters $ \alpha $ and $ \mu $ to vary as free parameters and fitted the velocity
dispersion data assuming $M_*/L_V=$ 0.5, 2, and 5 with the quality of the fits having an average $\chi^2\simeq 1.34 ,1.52, 1.27 $, respectively, which is the same as the approach of constant MOG parameters. The best-fitting values of  $ \alpha$ and $ \mu$ are larger than $ \alpha_{RC} $ and $ \mu_{RC} $ for almost all dSphs. The present results might mean that $ \alpha $ and $ \mu $ are not really universal constants in MOG as previously claimed and takes different values for different classes of objects. This impose two additional free functions to the theory which is problematic for MOG as a new law of physics.

In regards to the EFE, like many other authors, we assumed that in MOG the uniform external field does not affect the internal dynamics of a stellar system. However the MOG theory is linear in the weak field limit, there might be some screening effects in MOG. These effects (such as the chameleon effect or Vainshtein screening) arise from the hidden non-linearity in the field equations. Nobody has yet investigated this kind of non-linearity in MOG. Therefore, assuming the screening effect in MOG, one can expect to have EFE in MOG.  We leave this as our upcoming work to shed more light on the influence of screening effect on the non-linearity and possible EFE in MOG.


\section*{Acknowledgements}
We would like to thank the referee for constructive comments and suggestions. We would also like to thank Mahmood Roshan for fruitful and stimulating discussions of the manuscript.

\bsp \label{lastpage}

\end{document}